\documentclass{PoS}

\usepackage{amssymb}
\usepackage{gensymb}
\usepackage{graphicx}
\usepackage{amsmath}
\usepackage{wasysym}
\usepackage{verbatim}
\usepackage{wrapfig}
\usepackage{lineno}

\title{Dark Matter Annihilation and Decay Searches with the High Altitude Water Cherenkov (HAWC) Observatory}

\ShortTitle{HAWC Dark Matter Sensitivity}

\author{J. Patrick Harding$^a$ and \speaker{Brenda Dingus}$^a$ for the HAWC Collaboration$^b$ \\
        \llap{$^a$}Los Alamos National Laboratory, Los Alamos, NM, USA \\
        \llap{$^b$}For a complete author list, see \href{http://www.hawc-observatory.org/collaboration/icrc2015.php}{www.hawc-observatory.org/collaboration/icrc2015.php}.

        Email: \email{jpharding@lanl.gov}, \email{dingus@lanl.gov}}

\abstract{In order to observe annihilation and decay of dark matter,
  several types of potential sources should be considered. Some
  sources, such as dwarf galaxies, are expected to have very low
  astrophysical backgrounds but fairly small dark matter
  densities. Other sources, like the Galactic center, are expected to
  have larger densities of dark matter but also have more complicated
  backgrounds from other astrophysical sources. To search for
  signatures of dark matter, the large field-of-view of the HAWC
  detector, covering 2 sr at a time, especially enables searches
  from sources of dark matter annihilation and decay, which are
  extended over several degrees on the sky. With a sensitivity over
  2/3 of the sky, HAWC has the ability to probe a large fraction of
  the sky for the signals of TeV-mass dark matter. In particular, HAWC
  should be the most sensitive experiment to signals coming from dark
  matter with masses greater than 10-100 TeV. We present the HAWC
  sensitivity to annihilating and decaying dark matter signals for
  several likely sources of these signals.}

\FullConference{The 34th International Cosmic Ray Conference,\\
		30 July- 6 August, 2015\\
		The Hague, The Netherlands}

\begin{document}

\section{Introduction}

In recent years, the idea that dark matter is made up of weakly
interacting massive particles (WIMPs) has been particularly studied
(see ref.~\cite{Feng:2010gw} for more details). In this proceeding, we
consider the ability of the High Altitude Water Cherenkov (HAWC)
observatory to detect gamma rays associated with annihilation and
decay of multi-TeV-mass WIMP dark
matter~\cite{Abeysekara:2014ffg}. When dark matter annihilates or
decays, the emitted gamma-ray spectrum is expected to peak at roughly
an order-of-magnitude below the WIMP mass. This corresponds to a HAWC
sensitivity to WIMPs with masses between ~1\,TeV and ~1\,PeV. Many
dark matter gamma-ray sources are expected to have large angular
extent, so the HAWC Observatory is particularly sensitive to these
sources. Also, because HAWC will search over 2/3 of the sky, it will
be able to cross-correlate signals from many different types of dark
matter sources to improve its sensitivity and decrease systematic
uncertainties in the sources.

\section{Gamma Rays from Annihilating and Decaying WIMPs}\label{fluxes}

\subsection{Dark Matter Density Profiles}\label{DMprof}

From dark matter simulations, the density profile of dark matter
$\rho(r)$ falls off roughly as $r^{-2}$ away from the center of a dark
matter halo. Typically, the shape of this profile is chosen as either
an Navarro-Frenk-White (NFW) profile~\cite{Navarro:1996gj} or an
Einasto profile~\cite{Stadel:2008pn,Navarro:2008kc}, which have a
central cusp for small radii. The form we use for these profiles is
given in refs.~\cite{Abeysekara:2014ffg,Proper:2015}.

%
%
%
%
Other profiles which do not peak strongly toward the center of the
source are also possible~\cite{Burkert:1995yz}. However, even for dark
matter profiles which peak strongly toward the center of the source,
the dark matter halo for most sources is large. Therefore, most dark
matter signals are expected to give extended emission across the
sky. For the dark matter halo parameters of each source, we use the
parameters given in ref.~\cite{Abeysekara:2014ffg}.

\subsection{Gamma-ray Flux from Dark Matter Annihilation}

The gamma-ray flux from dark matter annihilation is a function of the
astrophysics of the source, including the dark matter mass-density
profile, $\rho(r)$, of the source as well the distance from the source
to the detector, $R$, and the angle it subtends on the sky,
$\Delta\Omega$. It also depends on the particle physics of the dark
matter model, including the WIMP mass $M_\chi$, the dark matter
annihilation rate $\langle\sigma_{\rm{A}}v\rangle$, and the spectrum
$\mathrm{d}N_\gamma/\mathrm{d}E$ of gamma rays from each WIMP
annihilation.

For dark matter annihilation, the gamma-ray flux depends on
the square of the dark matter density $\rho$ as
\begin{equation}\label{annJ}
\frac{\mathrm{d}F}{\mathrm{d}E}_{\rm annihilation}=\frac{\langle\sigma_{\rm{A}}v\rangle}{8\pi M_\chi^2}\frac{\mathrm{d}N_\gamma}{\mathrm{d}E}\int_{\Delta\Omega}\mathrm{d}\,\Omega\int \mathrm{d}\,x\ \rho^2(r_{\rm{gal}}(\theta,x))\enspace.
\end{equation}
To account for the total observed dark matter in the source, this
flux is integrated over the line-of-sight distance $x$, where the
distance from the center of the source is given by
\begin{equation}
r_{\rm{gal}}(\theta,x)=\sqrt{R^2-2xR\cos(\theta)+x^2}
\end{equation}
for an angle $\theta$ between the line-of-sight and the direction to
the source center. The gamma-ray spectrum $dN_\gamma/dE$ depends both
on the species of particle which the dark matter annihilates into and
the dark matter mass.

\subsubsection{Dark Matter Substructure}

Although the average shape of the dark matter density follows the
profiles described in section~\ref{DMprof}, dark
matter simulations show that dark matter halos also contain many
smaller subhalos, local overdensities of dark matter. Because the
gamma-ray flux from dark matter annihilation goes as the square of the
dark matter density, these denser regions actually contribute most of
the flux. This is particularly important for more massive objects,
such as galaxies and galaxy clusters, which have more mass and
therefore more massive and denser subhalos. Here, we consider the
substructure model of ref.~\cite{Sanchez-Conde:2013yxa}. In this
model, the dark matter flux is ``boosted'' by $\sim15$ for the M31
galaxy and by $\sim35$ for the Virgo cluster due to these subhalos. In
figures~\ref{HAWCM31} and ~\ref{HAWCVirgo}, we show the limits both
with and without this enhancement from substructure, for
comparison. Additionally, note that these values are very
conservative; in the substructure model of ref.~\cite{Han:2012uw}, for
example, the Virgo cluster can have its flux increased by 1000 times
over the flux of the smooth profile.

\subsubsection{Dark Matter Annihilation Cross-section}

Due to the uncertainty of the particle nature of dark matter, many
different models for the dark matter annihilation cross-section have
been hypothesized (for a review, see~\cite{Feng:2010gw}). However, one
of the most common models is a dark matter particle which is produced
thermally in the early universe. If such a particle annihilated in the
early universe, then the strength of the dark matter annihilation
cross-section should determine the amount of relic dark matter
observed today. Dark matter produced through such a mechanism should
have a cross-section between
$\langle\sigma_{\rm{A}}v\rangle\approx2.2\times10^{-26}\rm\,cm^3\,s^{-1}$
and
$\langle\sigma_{\rm{A}}v\rangle\approx4.4\times10^{-26}\rm\,cm^3\,s^{-1}$,
depending on the nature of the dark matter~\cite{Steigman:2012nb}. The
cross-section in the thermal dark matter model is largely independent
of the dark matter mass and is typically given as
$\langle\sigma_{\rm{A}}v\rangle\approx3\times10^{-26}\rm\,cm^3\,s^{-1}$,
which is the value we show in this paper.

Another common model consists of dark matter which interacts with
light dark-sector gauge bosons. In this model, called Sommerfeld
enhancement~\cite{Lattanzi:2008qa}, the exchange of these light bosons
can create a resonance between the dark matter particles, increasing
their cross-section by several orders of magnitude. The cross-section
in models with Sommerfeld enhancement is dependent on the relative
velocity of the dark matter particles as well as the mass and coupling
of the dark matter to the light bosons. For the high dark matter
masses considered in this paper, however, even standard model $W$ and
$Z$ bosons are light enough to give a Sommerfeld enhancement
effect. Any dark matter which is much heavier than the $W$-mass and
couples to standard model gauge bosons should produce this effect. For
the WIMP model we consider here, we assume a weak-scale coupling
between the dark matter and the $W$ boson, using the formalism of
ref.~\cite{Feng:2010zp}.

\subsection{Gamma-ray Flux from Dark Matter Decay}

In order for dark matter to survive with the present observed density,
dark matter particles must be fairly stable. However, it is possible
that dark matter, while long-lived compared to the age of the
universe, may have a finite lifetime. Even if the dark matter lifetime
were much longer than the age of the universe, the astrophysical
effects of the decay would be observable. Observations of
astrophysical signals, such as the neutrino flux observed by the
IceCube detector, could even be the first observations of PeV-mass
dark matter
decays~\cite{Aartsen:2013bka,Aartsen:2013jdh,Esmaili:2013gha,Bai:2013nga}.

The gamma-ray flux from dark matter decay is similar to the flux from
annihilating dark matter (equation~\ref{annJ}).  The flux for decay
depends on the dark matter decay lifetime $\tau_{\chi}$ instead of the
annihilation cross-section and the gamma-ray spectrum for each dark
matter decay.  Also, the flux depends on a single power of the dark matter
density $\rho$ as
\begin{equation}\label{decJ}
\frac{\mathrm{d}F}{\mathrm{d}E}_{\rm decay}=\frac{1}{4\pi\tau_\chi
  M_\chi}\frac{\mathrm{d}N_\gamma}{\mathrm{d}E}\int_{\Delta\Omega}\mathrm{d}\,\Omega\int
\mathrm{d}\,x\ \rho(r_{\rm{gal}}(\theta,x))\enspace.
\end{equation}
Because the flux is linearly proportional to the dark matter density,
dark matter substructures do not appreciably affect the flux from dark
matter decay. Also, decaying dark matter also gives more
spatially-extended emission than annihilation. The flux from decay is
much less sensitive to the shape of the dark matter density profile
than in annihilation. Therefore, the best limits on dark matter decay
are expected to come from the most massive objects in the universe,
including galaxies and galaxy clusters. Because the flux for dark
matter decay is inversely proportional to the dark matter lifetime
$\tau_{\chi}$, the limits on the dark matter lifetime are lower
limits, rather than the upper limits which are calculated for dark
matter annihilation.

\subsection{Gamma-ray Spectrum from Dark Matter Annihilation and Decay}
\begin{wrapfigure}{r}{0.5\textwidth}
\includegraphics[width=0.5\textwidth]{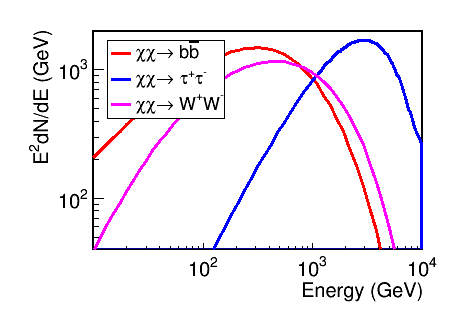}
\caption[Dark matter spectra]{The gamma-ray spectra for 10\,TeV dark
  matter particles annihilating into bottom quarks (red), tau leptons
  (blue), and $W$ bosons (magenta). The spectrum for the leptons is
  harder than for the quarks and bosons. All spectra cut off sharply
  at the dark matter mass.
\label{DMspectra}}
\end{wrapfigure}

Typically, the gamma-ray spectra are calculated assuming a 100 percent
branching fraction of the dark matter into a single final state. The
particles in this final state then decay to stable particles close to
the dark matter source, which produces gamma rays. For dark matter
annihilation, the maximum energy of a produced photon is the dark
matter mass, while for dark matter decay the maximum photon energy is
half the dark matter mass. Additionally, the expected dark matter
gamma-ray spectra from annihilation and decay are different from
typical astrophysical spectra. The gamma-ray spectra for 10\,TeV dark
matter particles annihilating into bottom quarks, tau leptons, and $W$
bosons are shown in figure~\ref{DMspectra}.

To calculate the dark matter gamma-ray spectra presented here, we use
the {\sc PYTHIA 6.4} code~\cite{Sjostrand:2006za}. This simulates
radiation from charged particles as well as the production of gammas
in decays of particles such as the $\pi^0$. We run this program
following the methods of ref.~\cite{Abazajian:2011ak} to calculate the
average spectrum $\mathrm{d}N_\gamma/\mathrm{d}E$ of gamma rays from each WIMP
annihilation or decay.

%

\section{HAWC Sensitivity to Dark Matter Annihilation and Decay}

\begin{figure*}
\begin{center}$
\begin{array}{c}
\begin{array}{cc}
\includegraphics[width=0.5\textwidth]{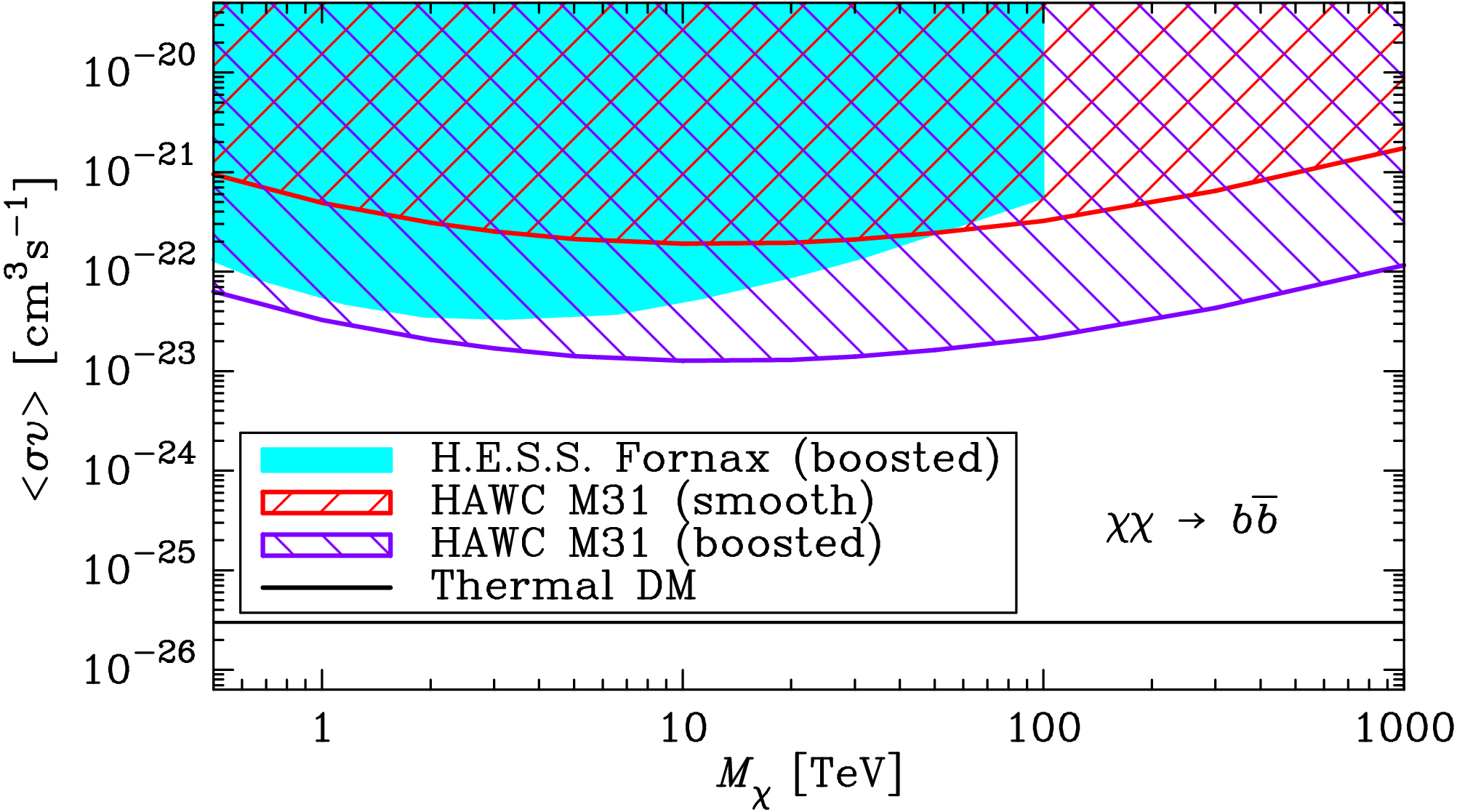} &
\includegraphics[width=0.5\textwidth]{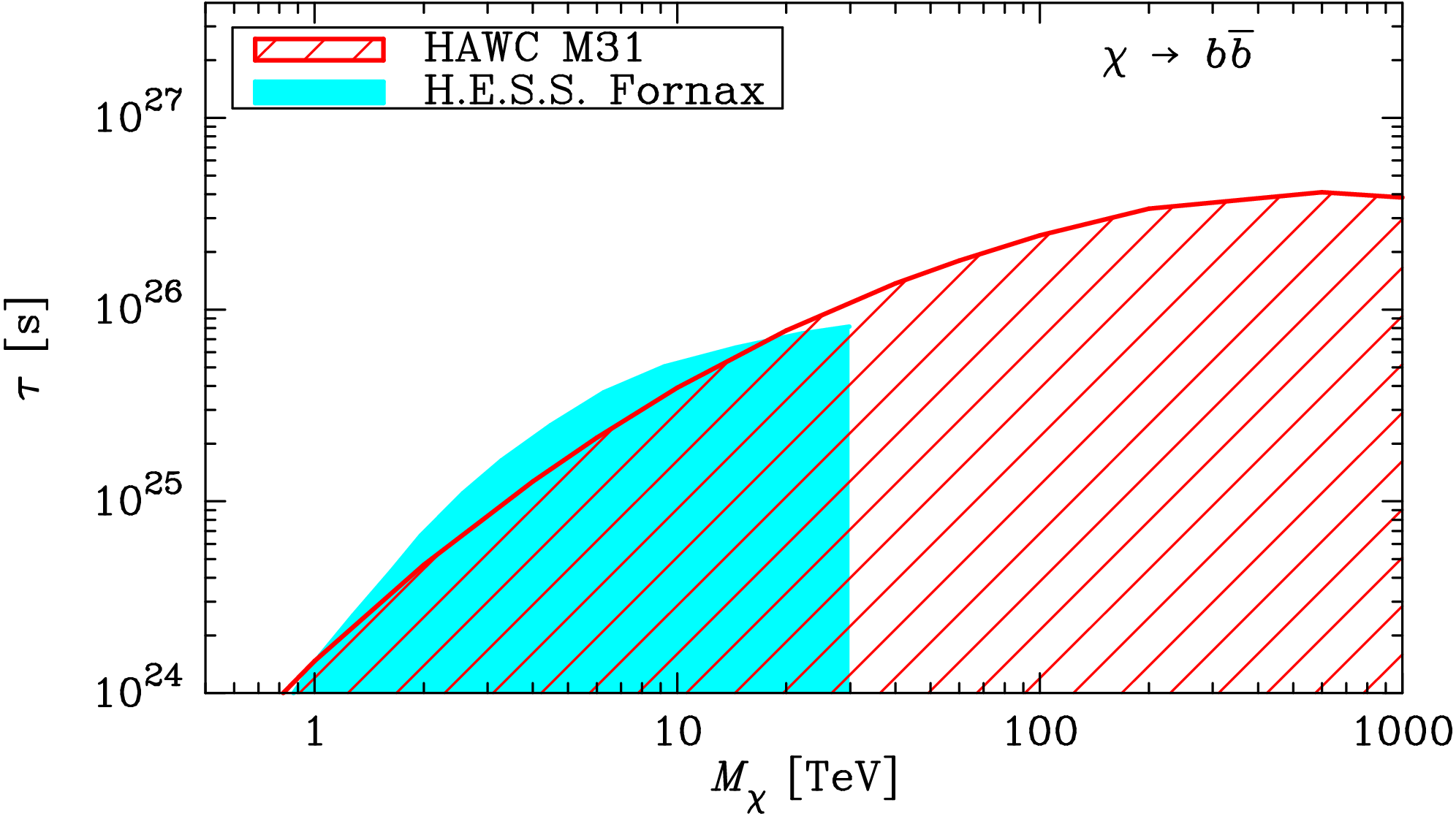}
\end{array}\\
\begin{array}{cc}
\includegraphics[width=0.5\textwidth]{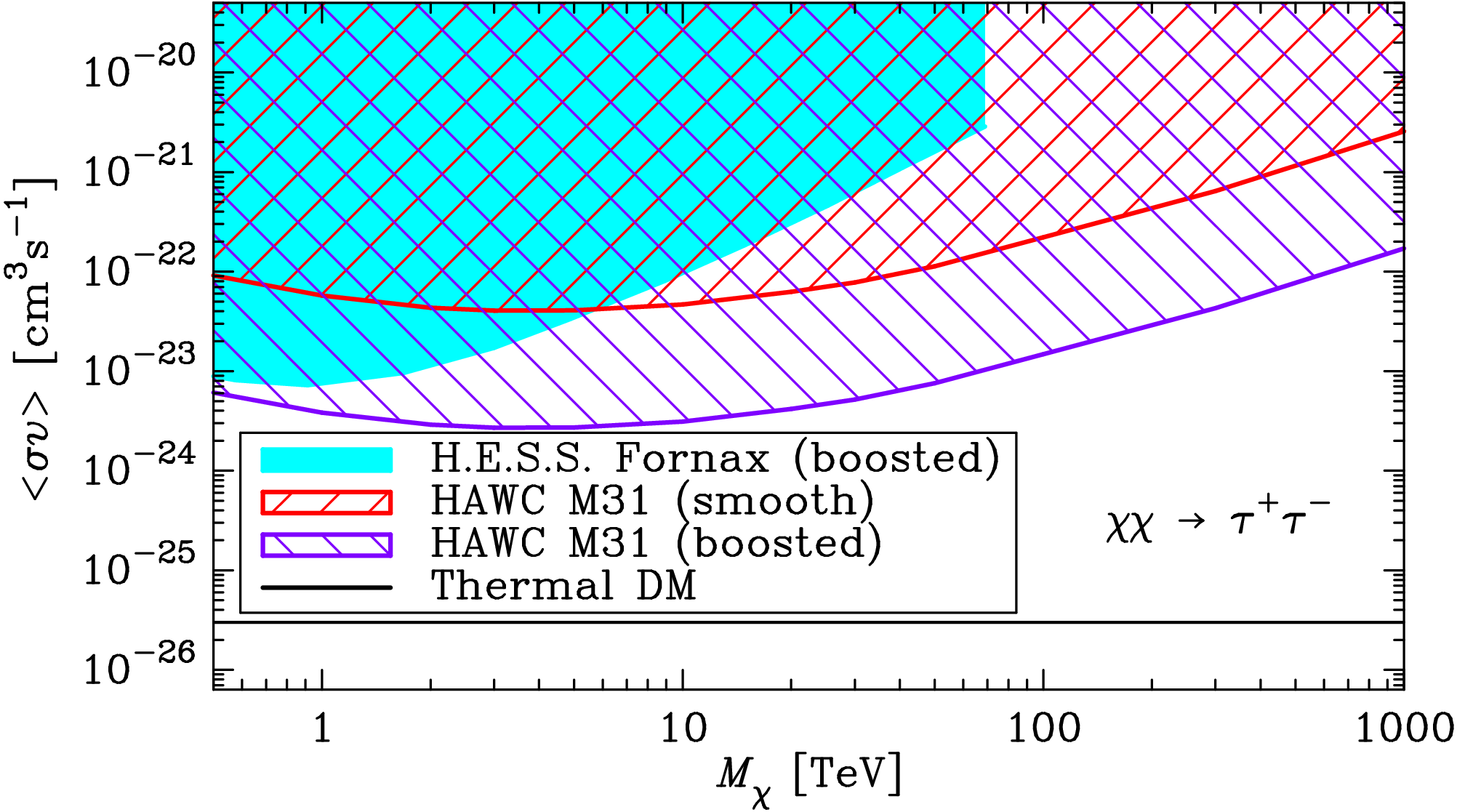} &
\includegraphics[width=0.5\textwidth]{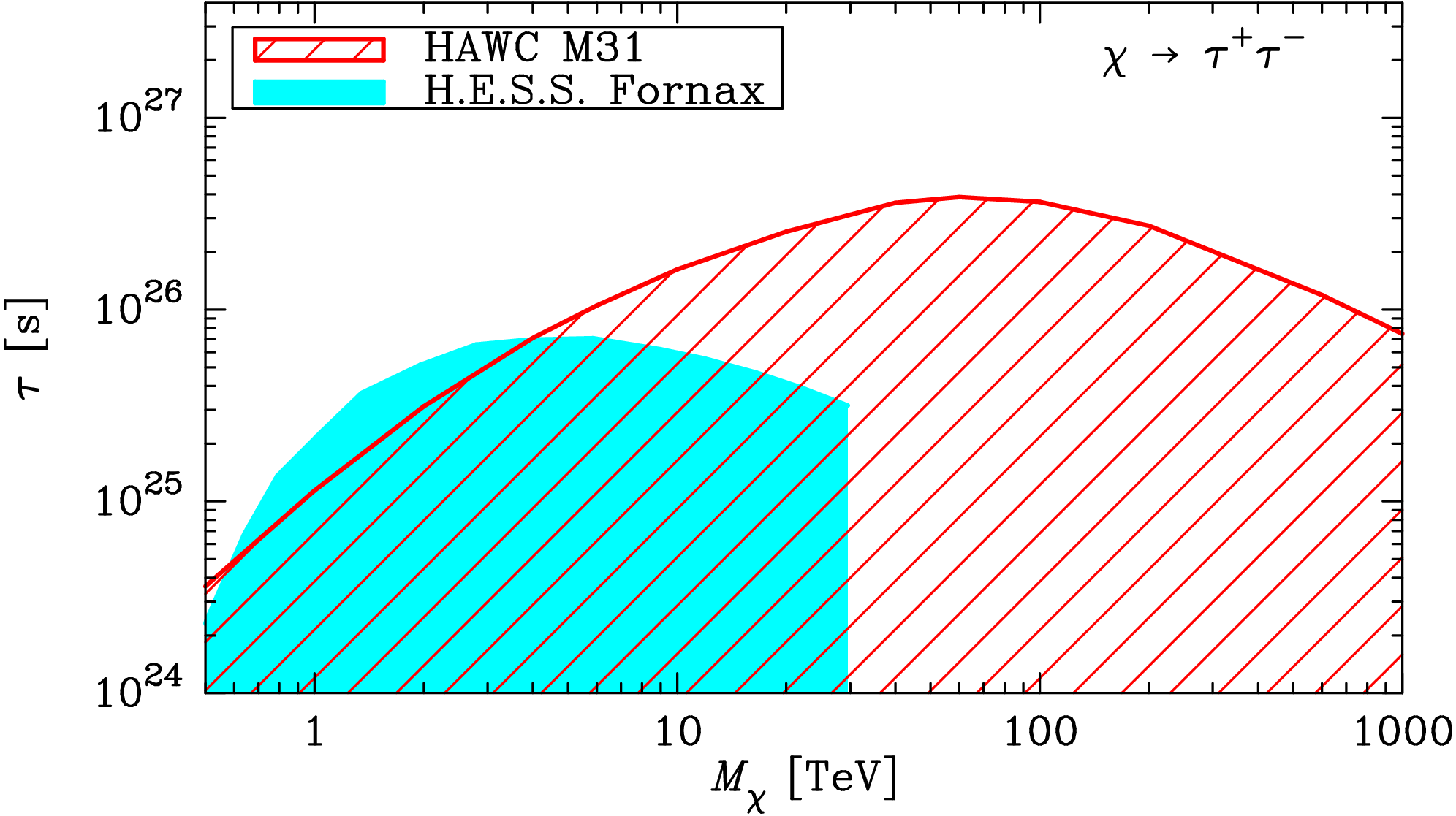}
\end{array}
\end{array}$
\end{center}
\caption[The projected dark matter limits from M31 for HAWC after five
  years.]{The projected dark matter limits from M31 for HAWC after
  five years, as a function of the dark matter particle mass. The
  plots on the left are the expected HAWC upper limits on the dark
  matter cross-section $\langle\sigma_{\rm{A}}v\rangle$ for dark
  matter which annihilates into bottom quarks (top) and tau leptons
  (bottom). The figures on the right are the expected HAWC lower
  limits on the dark matter lifetime $\tau_{\chi}$ for dark matter
  which decays into bottom quarks (top) and tau leptons (bottom). In
  the annihilation plots, the limits are shown both for just the
  smooth dark matter halo (red hatched) or with a substructure boost
  from ref.~\cite{Abramowski:2012au} (purple hatched). For comparison,
  the cyan regions show the measured limit from the
  H.E.S.S. observatory's observations of the Fornax
  cluster~\cite{Abramowski:2012au,2012PhRvD..86h3506C}, both for
  annihilating dark matter boosted using the substructure boost model
  of ref.~\cite{Sanchez-Conde:2013yxa} and for decaying dark matter,
  for comparison. The black line in the annihilation plots shows the
  expected cross-section for thermally-produced WIMP dark matter.  All
  limits are at 95\% CL.
\label{HAWCM31}}
\end{figure*}
\begin{figure*}
\begin{center}$
\begin{array}{c}
\begin{array}{cc}
\includegraphics[width=0.5\textwidth]{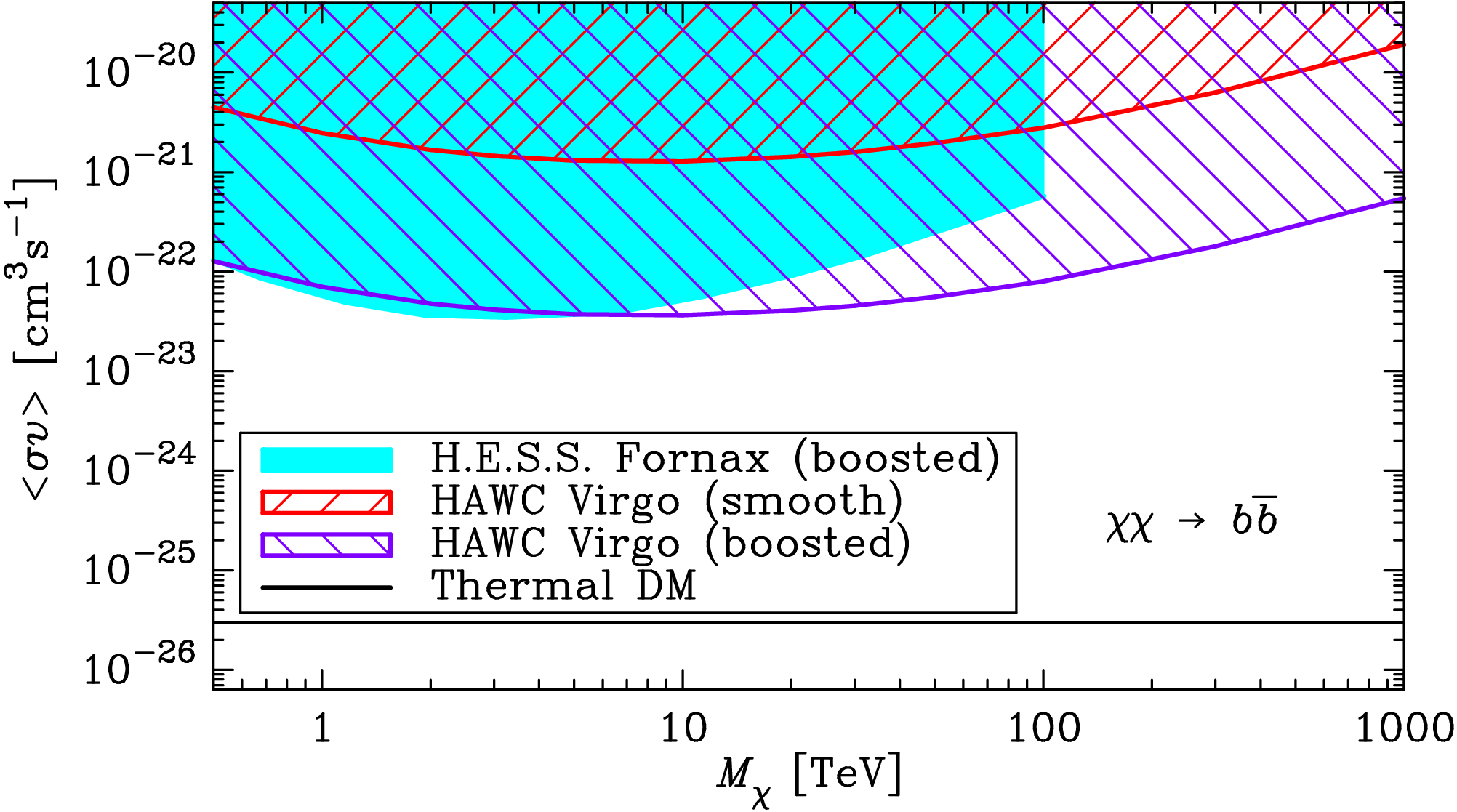} &
\includegraphics[width=0.5\textwidth]{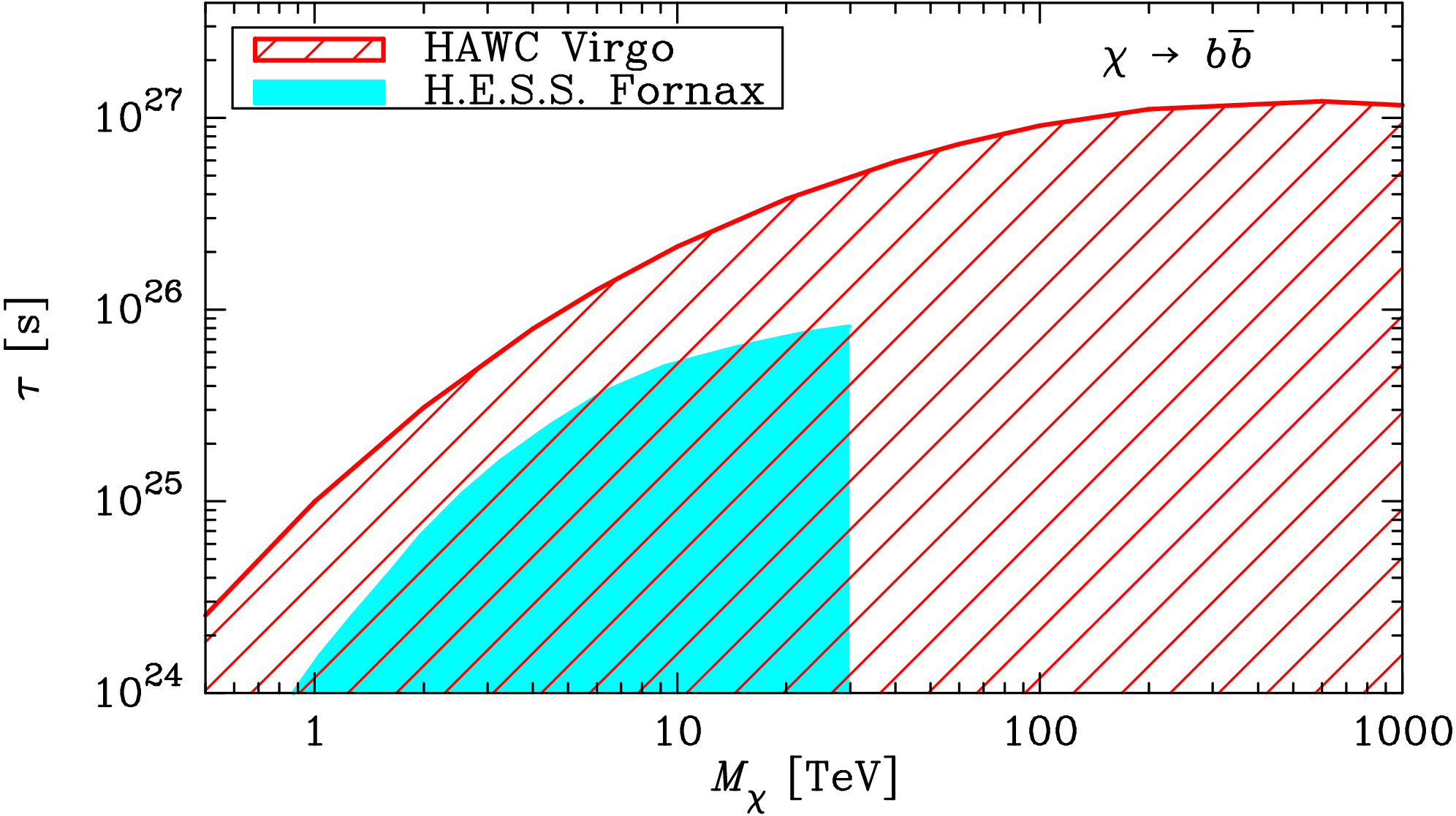}
\end{array}\\
\begin{array}{cc}
\includegraphics[width=0.5\textwidth]{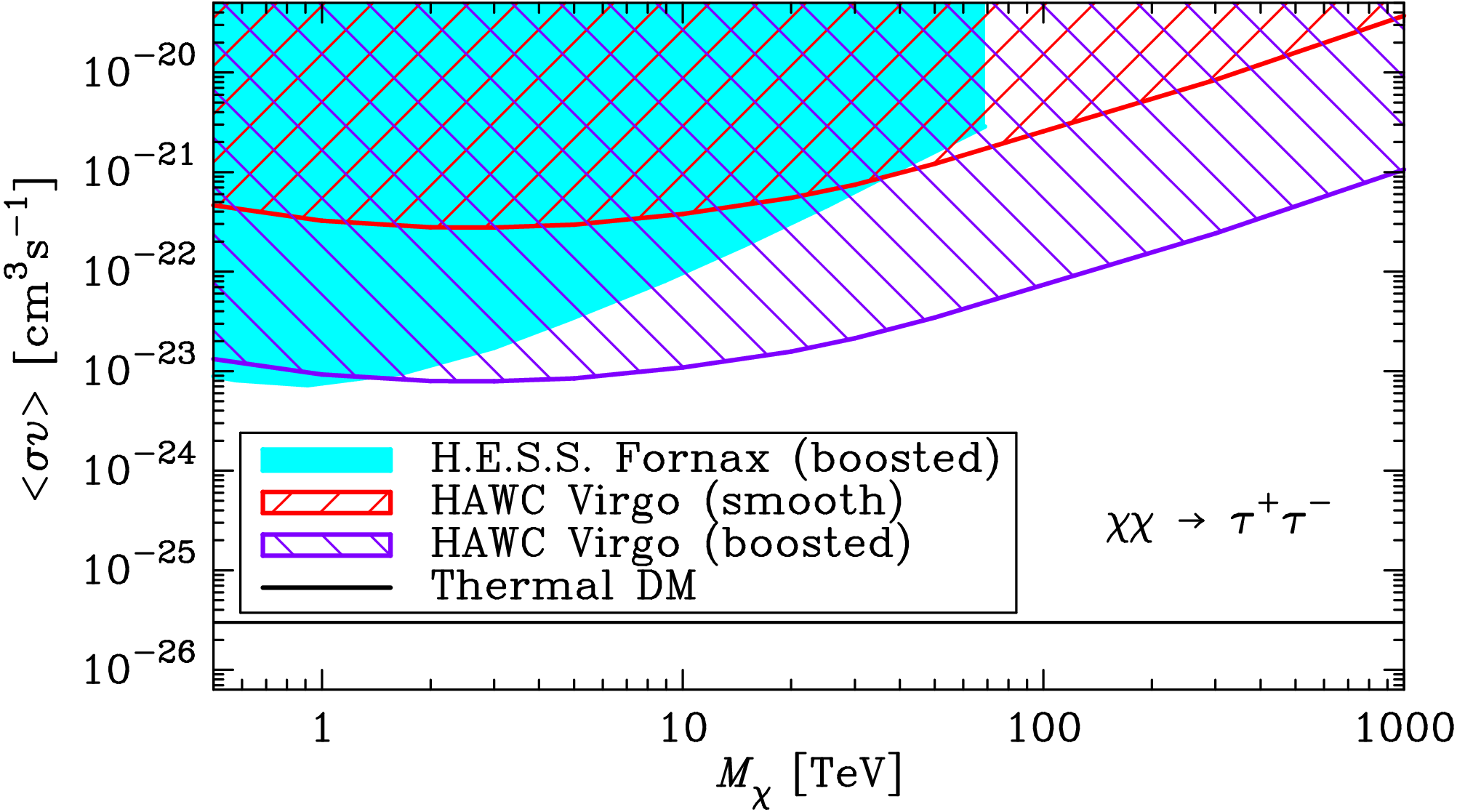} &
\includegraphics[width=0.5\textwidth]{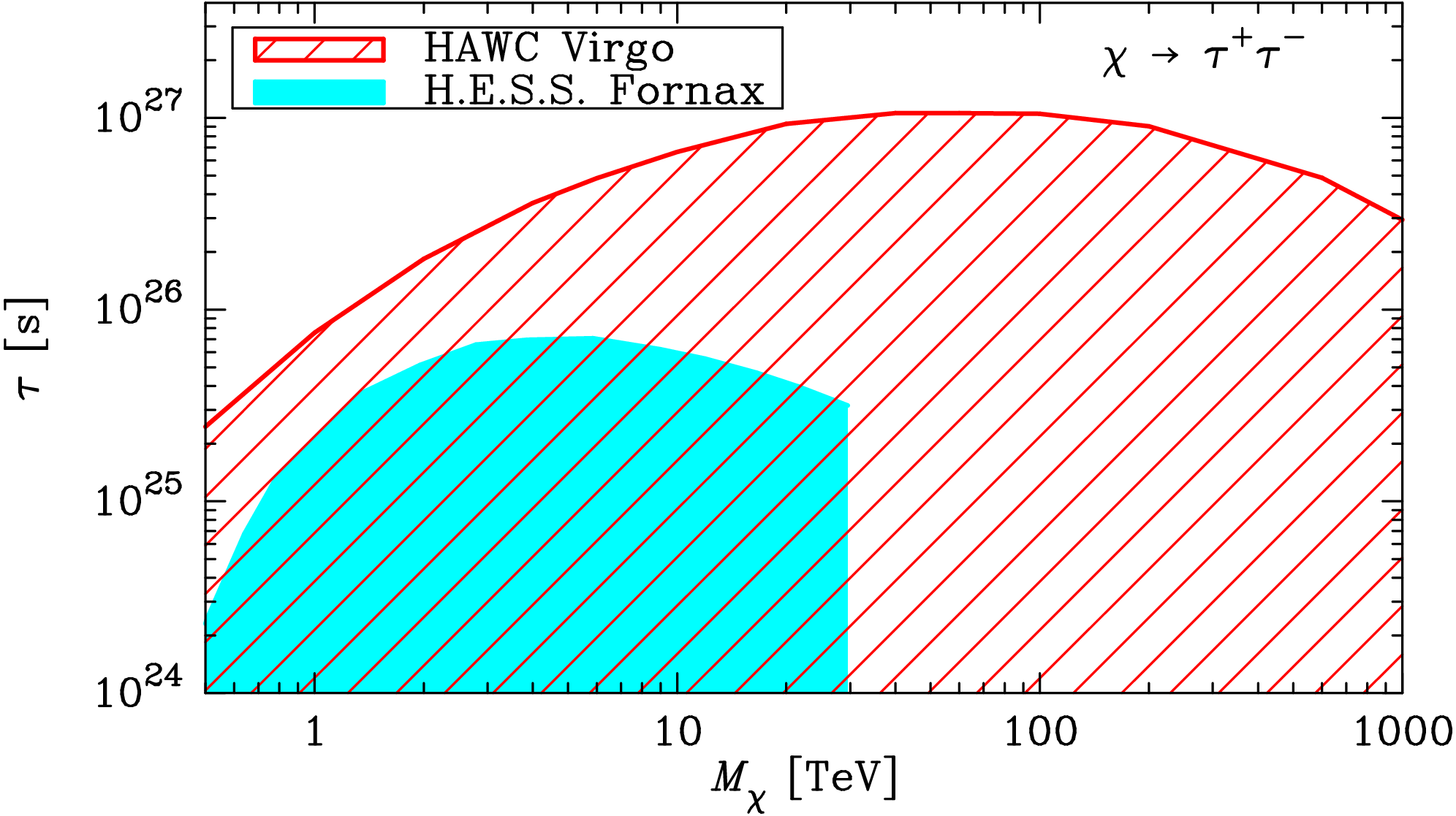}
\end{array}
\end{array}$
\end{center}
\caption[The projected dark matter limits from the Virgo Cluster for
  HAWC after five years.]{The projected dark matter limits from the
  Virgo Cluster for HAWC after five years, as a function of the dark
  matter particle mass. The figures follow the same format as those in
  figure~\ref{HAWCM31}.
\label{HAWCVirgo}}
\end{figure*}
\begin{figure*}
\begin{center}$
\begin{array}{c}
\begin{array}{cc}
\includegraphics[width=0.5\textwidth]{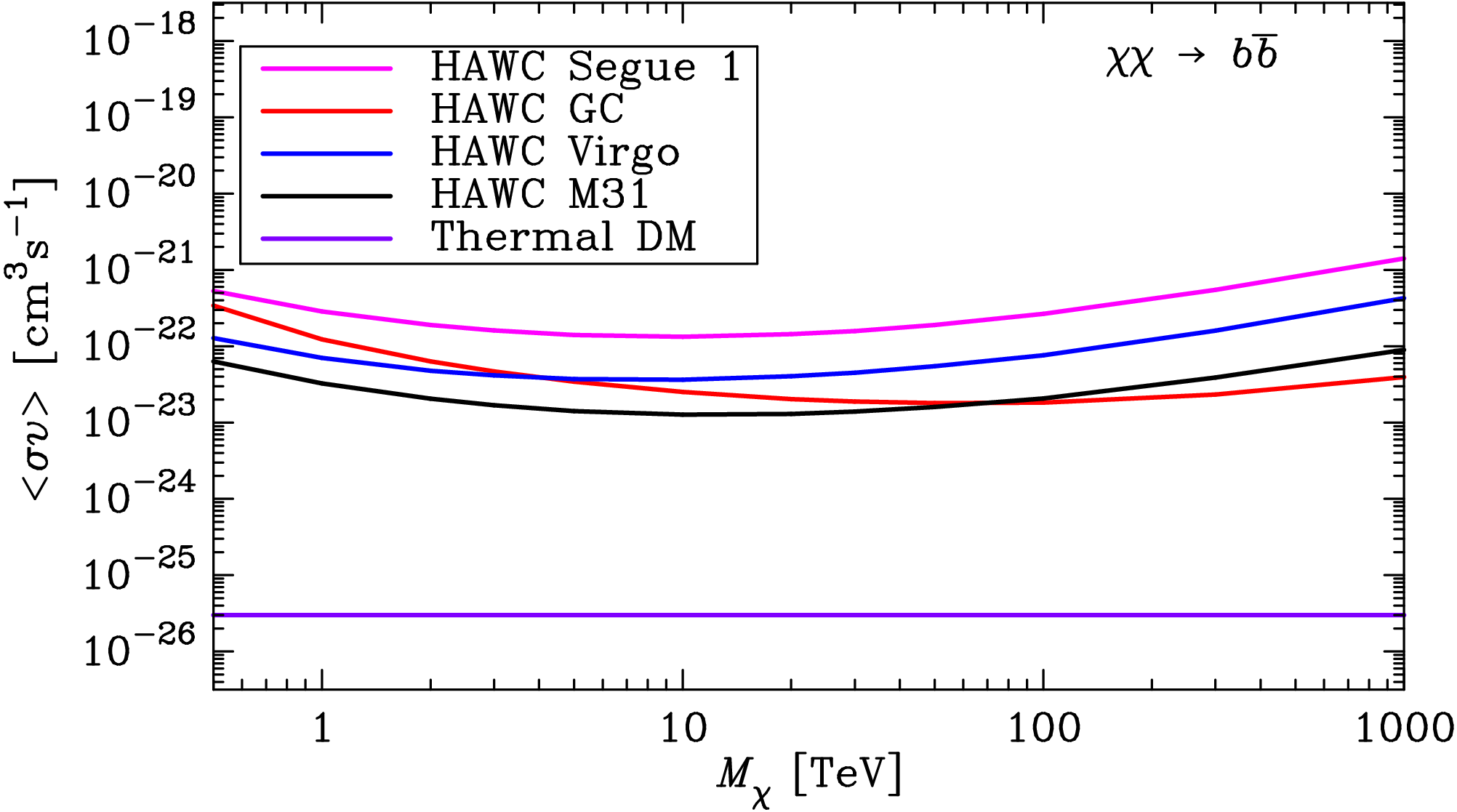} &
\includegraphics[width=0.5\textwidth]{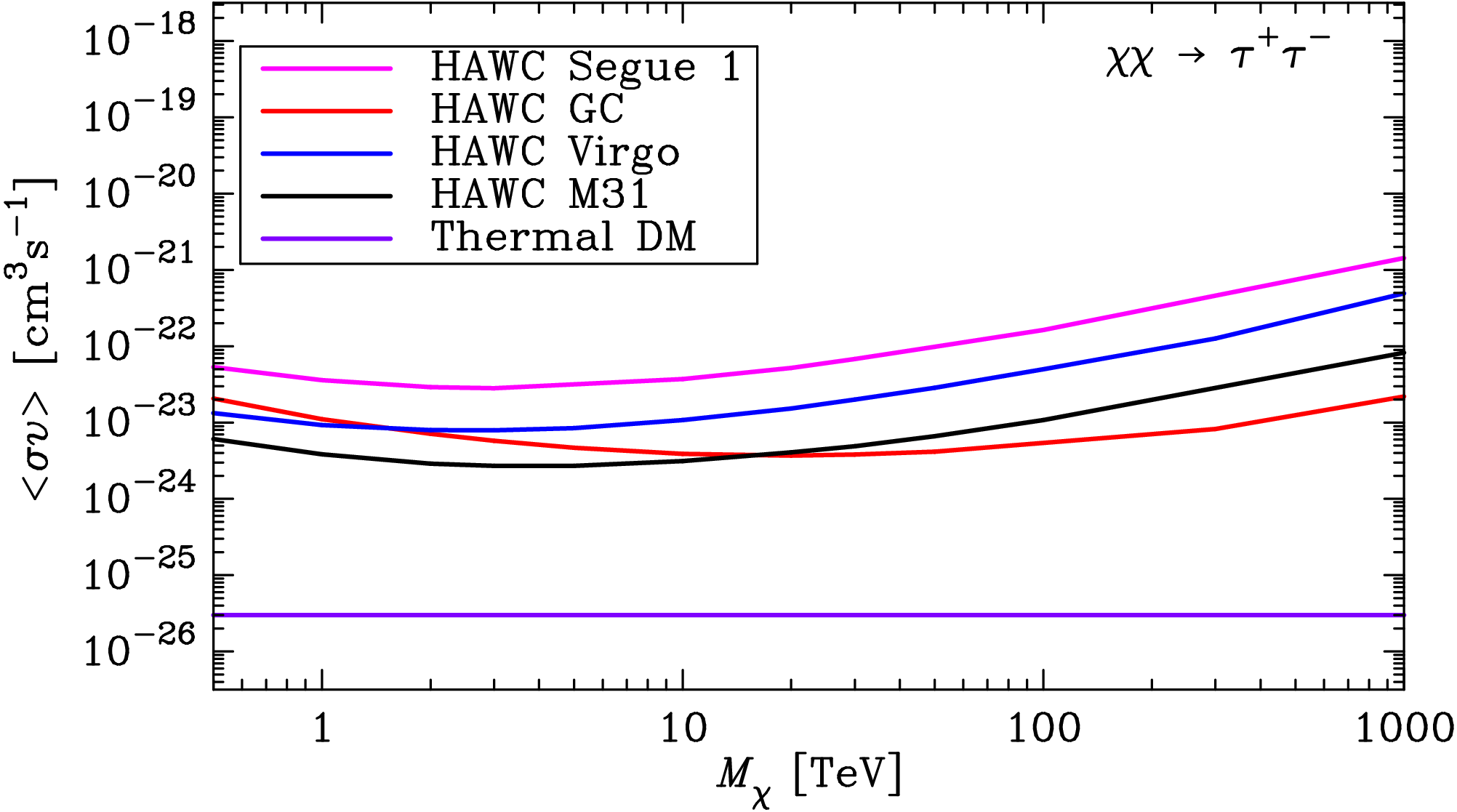}
\end{array}\\
\begin{array}{c}
\includegraphics[width=0.5\textwidth]{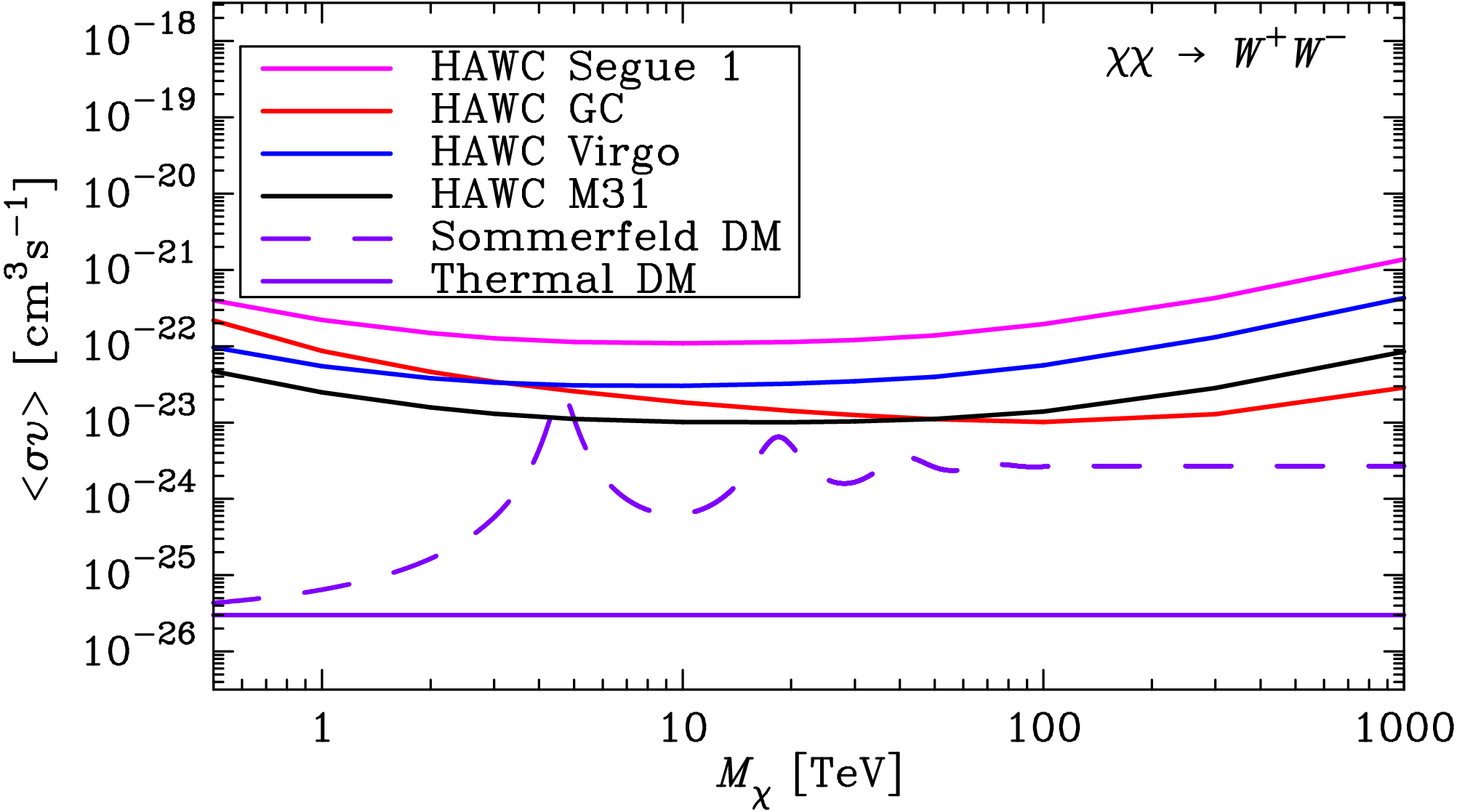}
\end{array}
\end{array}$
\end{center}
\caption[The projected dark matter limits from several sources for
  HAWC after five years.]{The projected dark matter upper limits from
  several sources for HAWC after five years, for the $b\bar{b}$,
  $\tau^+\tau^-$, and $W^+W^-$ dark matter annihilation channels.  The
  curves are for the Segue 1 dwarf galaxy (magenta), the Milky Way
  Galactic center (red), the Virgo cluster with substructure boost
  (blue), and the M31 galaxy with substructure boost (black). The
  substructure boosts follow the model of
  ref.~\cite{Sanchez-Conde:2013yxa}. The solid purple line shows the
  expected cross-section for thermally-produced WIMP dark matter. In
  the $W^+W^-$ figure, the dashed purple line indicates the
  cross-section expected for dark matter with natural Sommerfeld
  enhancement.  All limits are at 95\% CL.
\label{HAWClimits}}
\end{figure*}

Using the HAWC
simulation~\cite{Abeysekara:2013tza,Abeysekara:2014ffg}, we determined
the 5-year HAWC sensitivity to dark matter annihilation and decay for
several dark matter annihilation channels and a range of dark matter
masses. In figure~\ref{HAWCM31}, we show the HAWC sensitivity to dark
matter annihilation and decay in the M31 galaxy. For comparison, we
also show the measured limit from the H.E.S.S. observatory's
observations of the Fornax
cluster~\cite{Abramowski:2012au,2012PhRvD..86h3506C}, which is one of the strongest dark matter limits from galaxy clusters in the TeV mass range. In the
annihilation figures, the limit is shown both for just the smooth dark
matter density profile and for a substructure boost factor from the
model of ref.~\cite{Sanchez-Conde:2013yxa}. The H.E.S.S. Fornax limits
have an included substructure boost factor from the
ref.~\cite{Sanchez-Conde:2013yxa} model. For TeV-mass dark matter
annihilation, the limits from M31 should give HAWC's strongest
limits. For decaying dark matter, the HAWC
limits will place the strongest limits from a TeV experiment.

In figure~\ref{HAWCVirgo}, we show similar limits for the Virgo
cluster. The Virgo cluster limits from HAWC on dark matter
annihilation should be similar to the H.E.S.S. Fornax limit below 6
TeV (2 TeV) for the hadronic (leptonic) annihilation channel, with the
HAWC limits much more constraining at higher masses. Because the Virgo
cluster is the most massive dark matter object considered in this
study, the Virgo cluster will give the strongest HAWC limit on
decaying dark matter.

Figure~\ref{HAWClimits} shows the relative HAWC sensitivity to
annihilating dark matter for several sources and annihilation
channels. Although the substructure enhancement for the Virgo cluster
is larger than for M31, the annihilation limit from M31 will be
somewhat stronger than from the Virgo cluster. At the largest masses,
however, the Galactic center may provide the strongest limits from
HAWC. This is because the Galactic center peaks at $48\degree$ from
HAWC zenith, so many of the low-energy photons from the Galactic
center dark matter range out in the atmosphere and are not observable
by HAWC. The possibilities that contamination from the Galactic plane
or M87 could worsen these limits has been considered, but is expected
to affect the limits by less than 32\%~\cite{Abeysekara:2014ffg}. The
purple lines show the expected cross-section for the thermal dark
matter model. In the $W^+W^-$ channel plot, we also show the expected
cross-section for a Sommerfeld-enhanced dark matter model. For this
model, HAWC should be able to rule out a range of dark matter masses
after 5 years of observations.

\section{Conclusions}

With 5 years of observations, HAWC will be one of the most sensitive
experiments to TeV-mass dark matter annihilation and
decay~\cite{Abeysekara:2014ffg}. With its sensitivity to high
energies, HAWC will be able to observe signatures of dark matter with
masses above 1 TeV. Especially for extended dark matter sources,
including large sources such as M31 and the Virgo cluster, HAWC's
large field-of-view will enable observations of much of the dark
matter halo. The HAWC observations of the M31 galaxy, with its
expected substructure, will place the best HAWC constraint on models
of dark matter annihilation. For decaying dark matter, HAWC
observations of the Virgo cluster will provide the best TeV-scale
limits on the dark matter lifetime. Overall, the HAWC observatory
promises to open a new window on searches for dark matter annihilation
and decay.

\section*{Acknowledgments}
\footnotesize{
We acknowledge the support from: the US National Science Foundation (NSF);
the US Department of Energy Office of High-Energy Physics;
the Laboratory Directed Research and Development (LDRD) program of
Los Alamos National Laboratory; Consejo Nacional de Ciencia y Tecnolog\'{\i}a (CONACyT),
Mexico (grants 260378, 55155, 105666, 122331, 132197, 167281, 167733);
Red de F\'{\i}sica de Altas Energ\'{\i}as, Mexico;
DGAPA-UNAM (grants IG100414-3, IN108713,  IN121309, IN115409, IN111315);
VIEP-BUAP (grant 161-EXC-2011);
the University of Wisconsin Alumni Research Foundation;
the Institute of Geophysics, Planetary Physics, and Signatures at Los Alamos National Laboratory;
the Luc Binette Foundation UNAM Postdoctoral Fellowship program.
}

\bibliography{bibliography}

\end{document}